\documentclass[conference]{IEEEtran}
\IEEEoverridecommandlockouts
\usepackage{cite}
\usepackage{amsmath,amssymb,amsfonts}
\usepackage{algorithmic}
\usepackage{graphicx}
\usepackage{textcomp}
\usepackage{enumitem}
\usepackage{multirow}
\usepackage{amsmath}    
\usepackage{hyperref}   
\usepackage{diagbox}    

\usepackage{array}
\usepackage[table,xcdraw]{xcolor}

\usepackage{color, soul}
\def\BibTeX{{\rm B\kern-.05em{\sc i\kern-.025em b}\kern-.08em
    T\kern-.1667em\lower.7ex\hbox{E}\kern-.125emX}}
\usepackage{tcolorbox}  

\soulregister\cite7     
\soulregister\ref7
\soulregister\pageref7

\usepackage{listings}

\usepackage[table,xcdraw]{xcolor}

\definecolor{codegreen}{rgb}{0,0.6,0}
\definecolor{codegray}{rgb}{0.5,0.5,0.5}
\definecolor{codepurple}{rgb}{0.58,0,0.82}
\definecolor{backcolour}{rgb}{0.95,0.95,0.95}

\definecolor{pastelyellow}{rgb}{0.99, 0.99, 0.59}
\definecolor{lightgreen}{rgb}{0.56, 0.93, 0.56}
\definecolor{lightcoral}{rgb}{0.94, 0.5, 0.5}
\definecolor{azure(web)(azuremist)}{rgb}{0.94, 1.0, 1.0}

\lstdefinestyle{mystyle}{
  backgroundcolor=\color{backcolour},   commentstyle=\color{codegreen},
  keywordstyle=\color{magenta},
  numberstyle=\tiny\color{codegray},
  stringstyle=\color{codepurple},
  basicstyle=\ttfamily\scriptsize,
  breakatwhitespace=false,         
  breaklines=true,                 
  captionpos=b,                    
  keepspaces=true,                 
  numbers=left,                    
  numbersep=2pt,                  
  showspaces=false,                
  showstringspaces=false,
  showtabs=false,                  
  tabsize=2,
  frame=lines
}
\definecolor{light-gray}{gray}{0.95} 

\begin{document}

\title{A Qualitative Study on Using ChatGPT for Software Security: Perception vs. Practicality}


\author{\IEEEauthorblockN{
    M. Mehdi Kholoosi\IEEEauthorrefmark{1}\IEEEauthorrefmark{2}, 
    M. Ali Babar\IEEEauthorrefmark{1}\IEEEauthorrefmark{2},
    Roland Croft\IEEEauthorrefmark{1}\IEEEauthorrefmark{2}
    }
    \IEEEauthorblockA{\IEEEauthorrefmark{1} CREST - The Centre for Research on Engineering Software Technologies, The University of Adelaide, Adelaide, Australia}
    \IEEEauthorblockA{\IEEEauthorrefmark{2} Cyber Security Cooperative Research Centre, Australia}
    Emails: {\{mehdi.kholoosi, ali.babar, roland.croft}\}@adelaide.edu.au}

\maketitle

\begin{abstract}
Artificial Intelligence (AI) advancements have enabled the development of Large Language Models (LLMs) that can perform a variety of tasks with remarkable semantic understanding and accuracy. ChatGPT is one such LLM that has gained significant attention due to its impressive capabilities for assisting in various knowledge-intensive tasks. Due to the knowledge-intensive nature of engineering secure software, ChatGPT's assistance is expected to be explored for security-related tasks during the development/evolution of software. To gain an understanding of the potential of ChatGPT as an emerging technology for supporting software security, we adopted a two-fold approach. Initially, we performed an empirical study to analyse the perceptions of those who had explored the use of ChatGPT for security tasks and shared their views on Twitter. It was determined that security practitioners view ChatGPT as beneficial for various software security tasks, including vulnerability detection, information retrieval, and penetration testing. Secondly, we designed an experiment aimed at investigating the practicality of this technology when deployed as an oracle in real-world settings. In particular, we focused on vulnerability detection and qualitatively examined ChatGPT outputs for given prompts within this prominent software security task. Based on our analysis, responses from ChatGPT in this task are largely filled with generic security information and may not be appropriate for industry use. To prevent data leakage, we performed this analysis on a vulnerability dataset compiled after the OpenAI data cut-off date from real-world projects covering 40 distinct vulnerability types and 12 programming languages. We assert that the findings from this study would contribute to future research aimed at developing and evaluating LLMs dedicated to software security.
\end{abstract}


\section{Introduction}
\label{sec:introduction}

Software security remains one of the most significant challenges facing modern software development. With the rise of agile development practices, a focus is being placed on developer-centric security assurance \cite{rajapakse2022challenges}. However, studies have shown that developers fail to keep up with the ever-growing list of required security knowledge and expertise \cite{van2020schrodinger,mousavi2023detecting}. Secure coding alone requires developers to learn and use thousands of security-related patterns, practices, and tools. Some areas of software security, such as Software Vulnerability Management (SVM), need specialised knowledge and expertise for detecting, assessing, patching, and disclosing vulnerabilities \cite{mcgraw2008automated}. Consequently, developers tend to leverage recommended systems and/or social media platforms to gain and use knowledge about software security \cite{rajapakse2022challenges}. 

Chatbots, i.e., conversational agents, provide effective interfaces for enhanced information retrieval and task assistance \cite{frazier2022investigating}. Natural language comprehension offers a potentially game-changing interface facilitating human and computer interaction. Natural language can greatly improve the efficiency in which we perform complicated tasks and use complex technologies by allowing us to operate them with simple natural language queries. 

Like many other domains, such as education and marketing \cite{adamopoulou2020overview}, chatbots are gaining significant traction in the software development domain for obtaining knowledge/information related to specific software development tasks, such as API usages \cite{erlenhov2020empirical}. However, the use of chatbots for software security is still in its infancy stage \cite{tony2022conversational} as current software security chatbots use simple rule-based technologies with limited capabilities \cite{tony2022conversational}. This situation is expected to change as the advancements in Large Language Models (LLMs) enable much more powerful and capable chatbots, e.g., ChatGPT. 

ChatGPT is an Artificial Intelligence (AI) chatbot developed by OpenAI\footnote{https://openai.com/blog/chatgpt/} created using GPT (Generative Pretrained Transformer)\cite{brown2020language}, a state-of-the-art LLM. These technical advancements allow ChatGPT to surpass traditional chatbots through its capability to provide answers and solutions to complex queries
\cite{haque2022think}. 
Training and fine-tuning an LLM with billions of parameters (e.g., 175 billion in GPT-3\cite{brown2020language} and 1,700 billion in GPT-4\cite{openai2023gpt4}) presents considerable challenges due to the extensive computational resources required \cite{thapa2022transformer}. Since ChatGPT was released to the public, this limitation has been addressed, and many users were able to experience LLMs for the first time. Whilst ChatGPT was not specifically developed for the software domain, it still capture intrinsic knowledge that enables software security tasking to a certain degree \cite{liu2023no, sobania2023analysis, fu2023chatgpt}. We believe exploring the capabilities and limits of this intrinsic knowledge is extremely beneficial for end-users as most organisations and practitioners lack the required resources to teach domain-specific knowledge to LLMs. Hence, we intend to investigate ChatGPT's intrinsic knowledge in software security to help enable the use of chatbots in this area in the future.

We conducted a multifaceted exploration of perceptions and practicality to investigate ChatGPT's potential for supporting software security tasks. First, we analysed public discussions of ChatGPT users regarding software security to gain insights into the successes and failures of LLM-based chatbots for secure development. In the subsequent phase of our study, we focused on vulnerability detection as it was the most anticipated security task among practitioners, according to the first step of our analysis. 
Through qualitative analysis of ChatGPT's responses, we obtained an overview of the types and manners in which information (e.g., descriptions and assessments) is provided and examined the responses for potential practicality issues. Using this two-staged comprehensive assessment, we uncovered not only the current capabilities and shortcomings of ChatGPT in the context of software security but also demonstrated how practitioners' expectations differ from the practical application of this technology in the real world. We assert that the findings from this study are expected to guide future endeavours that aim to develop specialised LLMs for security-related tasks.
\section{Related Work}
\label{sec:relatedwork}


\subsection{AI-Based Chatbots for Software Security.}
Chatbots have been demonstrated to be suitable for software engineering tasks such as information retrieval \cite{frazier2022investigating} and API usage \cite{tian2017apibot}. However, the use of chatbots for software security is limited. To date, the SKF-Chatbot (Secure Knowledge Framework) is one of the leading software security chatbots, which was developed by the OWASP Foundation\footnote{https://owasp.org/} to provide access to Software Vulnerability (SV) information via a chat interface. However, its capabilities are limited due to its lack of contextual understanding \cite{tony2022conversational}. Recently, AI-based chatbots have been able to overcome these limitations via the LLMs that power them. ChatGPT has thus far demonstrated remarkable capabilities for semantic comprehension and providing tailored solutions for given tasks \cite{haque2022think}. 

\subsection{Vulnerability Detection Using Transformer-based Language Models.}
Prior studies have explored the use of transformer-based language models to identify software vulnerabilities using a variety of fine-tuning methodologies. For example, Thapa et al. \cite{thapa2022transformer} evaluated  transformer-based language models (i.e., BERTBase, GPT-2 Base) against recurrent neural network (RNN)-based models (i.e., BiLSTM, BiGRU) for vulnerability detection in software datasets featuring C/C++ source code. According to their findings, transformer-based language models outperformed RNN-based models on all evaluation metrics. In another study, Kalouptsoglou et al. \cite{kehagiasempirical} fine-tuned various transformer-based models (i.e., BERT variants, GPT-2, BART) and carried out a comparative analysis to determine which of these models is the most appropriate for vulnerability detection.
Furthermore, Chan et al. \cite{chan2023transformer} created a vulnerability detection model targeted at detecting vulnerabilities in incomplete code snippets. They leveraged common learning approaches such as fine-tuning on three pre-trained LLMs, namely CodeBERT, code-davinci-002, and text-davinci-003. 

\subsection{Vulnerability Detection Using ChatGPT.}
Several research efforts have explored the efficacy of ChatGPT in vulnerability detection. 
Chen et al. \cite{chen2023chatgpt} exclusively probed ChatGPT's capability in identifying smart contract vulnerabilities. Szabó et al. \cite{szabo2023new} focused on identifying vulnerabilities associated with CWE-653 (Improper Isolation or Compartmentalization) using multiple GPT models. Ozturk et al. \cite{ozturk2023new} assessed ChatGPT's effectiveness in detecting the top 10 OWASP vulnerability categories in web applications. Zhang et al., \cite{zhang2023prompt} through comprehensive testing on two Java and C/C++ 
vulnerability datasets, demonstrated that prompt-enhanced approaches could strengthen ChatGPT's vulnerability detection capability.
In a recent study, Fu et al.\cite{fu2023chatgpt} compared the capabilities of ChatGPT (i.e., GPT-3.5 and GPT-4) against three other LLMs (e.g., CodeBERT\cite{feng2020codebert}) across various vulnerability tasks. The LLMs used in this study were fine-tuned explicitly for code-related tasks.
\newline
There are two main ways in which our study differs from prior studies:
\begin{enumerate}

    \item 
    Prior research has primarily focused on assessing \cite{ozturk2023new,thapa2022transformer,kehagiasempirical,fu2023chatgpt} and enhancing ChatGPT's vulnerability detection performance through prompt engineering methodologies \cite{ullah2023can,nong2024chain,zhang2023prompt} or the integration of supplementary elements \cite{mathews2024llbezpeky} within the LLM pipeline.
    We instead strive to focus more on insights through qualitative analysis of existing real-world uses and outputs. This shift aims to illuminate the disparity between practitioners' perceived value and the practicality of this technology in real-world settings. We provide further details in Section \ref{sec:results}.
    \item Prior studies that utilised datasets from existing literature \cite{kehagiasempirical, zhang2023prompt, chen2023chatgpt, fu2023chatgpt} or open-source projects \cite{thapa2022transformer, chan2023transformer, szabo2023new} ignored the possibility of data leakage (i.e., overlap between training and test data). This factor could artificially inflate the reported performance of AI models' (e.g., ChatGPT) \cite{croft2023data,gennari2024considerations,sallou2023breaking}. Consequently, we curated a high-quality vulnerability dataset tailored to our evaluation criteria from unseen data. Section \ref{sec:dataset-vulnerability} discusses this in more detail.
    
\end{enumerate}

\section{Research Design}
\label{sec:method}
\begin{figure}[h]
  \centering
  \includegraphics[width=0.98\linewidth]{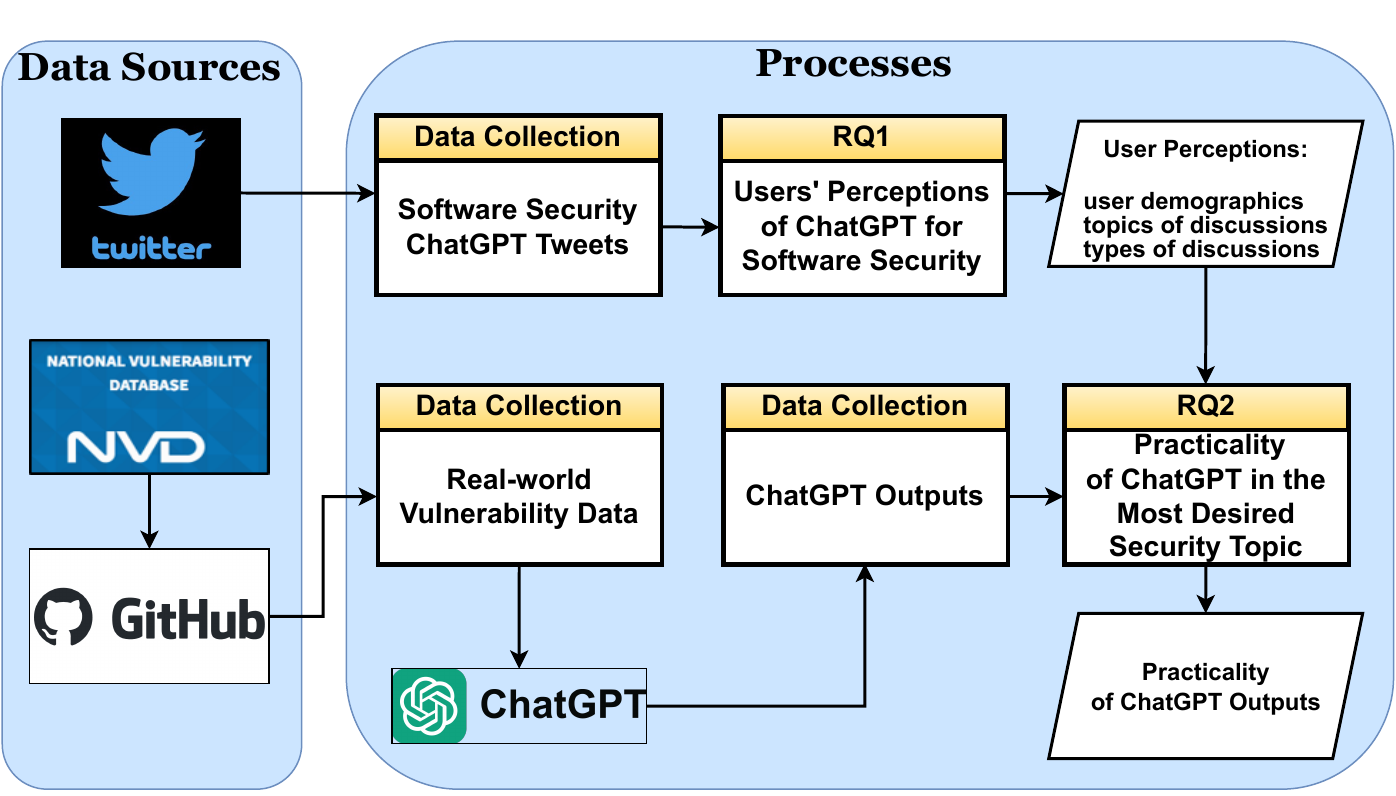}
  \caption{The overall study design.}
  \label{fig:overview}
\end{figure}

Initially, we present our research questions in Section \ref{sec:Research Questions}. This is followed by a detailed explanation of the data collection methodology in Section \ref{sec:data collection}. Lastly, Section \ref{sec:qualitative analysis} delves into the qualitative analysis processes.
Figure \ref{fig:overview} displays the overall workflow that we used to conduct this study.

\subsection{Research Questions.}
\label{sec:Research Questions}
The following Research Questions (RQs) motivated our empirical study: 
\begin{itemize}
    \item \textbf{RQ1: What are users' perceptions of using ChatGPT for software security?} We examine user demographics, topics, and types of discussions to determine how ChatGPT is being applied by users and its potential strengths and weaknesses.


    

    \item \textbf{RQ2: How practical are ChatGPT's outputs for the most desired software security task?} We perform a qualitative analysis of ChatGPT's outputs for the most discussed topic of RQ1 to ascertain the practicality of this technology in real-world settings.
\end{itemize}
\subsection{Data Collection.}
\label{sec:data collection}
\subsubsection{\textbf{Collecting Twitter Data and Curating a Sample}} For the first data source, we selected the popular social media site, Twitter\footnote{https://twitter.com}, for our investigation due to its active use by developers to discuss emerging technologies \cite{singer2014software} and its use in former research for uncovering public sentiments \cite{dong2021review, bian2016mining}.
We used the Twitter API to collect English language tweets that matched the following keyword search query: \texttt{ChatGPT AND (security OR secure OR insecure OR vulnerability OR vulnerabilities)}. We collected tweets that were posted between December 1, 2022, and February 28, 2023, the first three months of ChatGPT's public release. 
In total, we collected 7716 tweets.



To enable qualitative analysis of our data, we curated a sample of posts for manual examination. To ensure that our data sample was of sufficiently high quality, we selected the 700 most liked tweets (approximately 10\% of the collected tweets) from our dataset. Twitter likes are given by users who appreciate or agree with a tweet. Hence, highly liked tweets are influential posts that reflect the viewpoints of a larger audience and represent broader public sentiments. We label these tweets T1-T700.

\subsubsection{\textbf{Collecting Vulnerability Data and Curating a Sample}}
\label{sec:dataset-vulnerability}
To perform the required analysis for RQ2, 
we needed to test ChatGPT's capabilities in real-world settings. This purpose was achieved by using the National Vulnerability Database (NVD) \cite{NVD}, an established repository for SV management data maintained by the National Institute of Standards and Technology (NIST). This database holds an exhaustive array of information on identified SVs in the wild. 

As we planned to do an in-depth manual analysis for RQ2, 
we decided to work with a subset of 70 SVs.
This sample size was of statistical significance, providing 90\% confidence and an error margin of 10\% \cite{cochran2007sampling}. Previous researchers who needed to perform manual analysis over SVs also chose a similar sample size \cite{croft2023data,le2024latent}. 
It is imperative to note that we considered the issue of data leakage (i.e., contamination of training and test data) in our data selection for RQ2. 
It is well known that ChatGPT has been trained using a great deal of publicly available data. Since NVD is a public database, it is possible that ChatGPT encountered NVD data during its training phase. Thus, to prevent data leakage, we only included SVs in our dataset with a published date (i.e., the official date on which a vulnerability is added to the NVD) after Sep 2021 (the knowledge cutoff date for ChatGPT at the time of conducting this study). Any newly discovered SVs after this date have not been seen by ChatGPT during training.

Additionally, three other factors were considered when curating the random sample of 70 SVs. First, as the fixed code was required for validation purposes in our analysis, the SV report on NVD must be directly linked to the GitHub commit that addresses the SV (i.e., fixing commit). The \textit{References} section of each SV in NVD usually includes a link to the fixing commit. The second consideration is that the fixing commit should be focused and not tangled (i.e., a commit that consists of multiple unrelated changes). 
The last considered factor is that the vulnerability should be contained within a single function. The last two factors are because we planned to perform all analysis in the following steps manually, and tracing complicated commits is an error-prone process.

After incorporating the aforementioned factors into our SV selection criteria, we ended up with a dataset (containing 70 SVs) that covered 40 distinct CWEs (Common Weakness Enumeration) \cite{CWE}, and its affected products (i.e., products and platforms considered vulnerable) were written in 12 different programming languages. 
By utilising the provided fixing commits, we carefully collected the entire vulnerable and fixed (i.e., patched) versions of functions from the relevant GitHub repositories. The average number of lines of code among our 70 collected vulnerable functions was 32.

\subsubsection{\textbf{Collecting ChatGPT Outputs}} \label{sec:dataset-ChatGPT}
In this step, ChatGPT was fed the 70 collected vulnerable functions as input and asked about potential SVs within them. The GPT-4 model has demonstrated superior performance compared to the GPT-3 model in various benchmarks \cite{openai2023gpt4}. Considering the importance of delivering high-quality and up-to-date research insights in the critical domain of software security, all ChatGPT-related analysis in this paper were performed using the GPT-4 model. 
Prior research has shown that tailored prompts are essential to taking full advantage of LLM's capabilities in various software engineering tasks\cite{white2023chatgpt,feng2023prompting,zhang2023prompt}. 
Following their prompt designs, for each of the SVs in our dataset, we utilised the following prompt template in our conversation with ChatGPT and carefully saved its outputs 
for further analysis. As ChatGPT remembers information from previous inputs, each query was made in a new chat. 


\begin{tcolorbox}[colback=blue!5!white,colframe=blue!75!black,title= Prompt Template]
  \textbf{Role:} You are a software security expert.
  
  \textbf{Instruction:} Please analyze the following code snippet for potential security vulnerabilities. Provide a detailed explanation of the issues you find.
  
  \textbf{Context:} $<code-snippet>$
\end{tcolorbox}

\subsection{Qualitative Analysis.}
\label{sec:qualitative analysis}
\subsubsection{\textbf{Qualitative Analysis of Sampled Tweets}}
\label{sec:tweet-analysis}
To understand user perceptions of ChatGPT for software security (RQ1), we qualitatively analysed our sample of 
700 tweets using thematic analysis \cite{cruzes2011recommended}. This was an iterative and reflective process that we conducted over several phases. 

We started by immersing ourselves in the data by carefully reading each tweet and all surrounding data: other tweets in the thread, replies, attached media, linked articles, and associated user profiles. The first and third authors conducted this pilot data labelling process collaboratively for 30 tweets, and it typically took 15 to 20 minutes per tweet. During this process, we posed potential characteristics of the discussion that address RQ1: user demographics, topics and types of discussion. To determine the values of these characteristics, we performed open coding \cite{cruzes2011recommended} to classify the common discussion points. We 
collaboratively identified \textit{keypoints} as short summaries for each tweet. We then revised these into \textit{themes}, which provide a one or two word description of the common topics. We conducted a second round of pilot data labeling on an additional batch of 30 randomly selected tweets to ensure thorough coverage of potential themes. In this iterative process, no new themes emerged, indicating that thematic saturation had been reached. 

Once we were satisfied with the identified themes, two authors independently analysed and categorised the remaining 640 
tweets in the sample. They then compared the annotated data to ensure data coding was conducted consistently. The two raters achieved a Cohen Kappa value of 0.813 \cite{cohen1960}, which implies a high level of agreement. Any remaining disagreements between the first and third authors were resolved through discussion with the second author. During the labelling process, we excluded any tweets from our sample that were unrelated (434 tweets) to the use of ChatGPT for software security due to falsely matched keywords, e.g., tweets discussing the impact of ChatGPT on job `security'.
Ultimately, 266 tweets were included in the sample.

We also wanted to gain an understanding of general user sentiments toward ChatGPT's use for software security. E.g., whether the author of a tweet had a positive experience using it or was optimistic about its potential applications. We experimented with using the NLTK sentiment analysis tool \cite{bird2009natural} to classify our data automatically. However, we found this tool to be inaccurate from a manual review of 30 sample tweets. This is because security keywords such as \textit{vulnerability} or \textit{security} were often falsely interpreted as indicating negative sentiments. Furthermore, a tweet may have an overall positive sentiment but a negative opinion of ChatGPT, and vice versa. For example, \textit{``This article does a really good job of highlighting the weakness of ChatGPT.''} Hence, we manually classified user sentiments of ChatGPT during qualitative analysis. For each tweet, we labelled the sentiment as either positive, negative, or neutral based on our interpretation of the user's opinion or experience of ChatGPT capabilities for software security. We reserved the neutral sentiment label for tweets that discussed positive and negative points or were ambiguous to identify. The nature of the sentiment of a tweet depends on the type of discussion (e.g., experiences vs. speculation). Hence, we consider the types of discussion whilst performing sentiment analysis, which we elaborate further on in Section \ref{sec:results-RQ1}.


\subsubsection{\textbf{Qualitative Analysis of Sampled Vulnerabilities}}
\label{sec:vuln-analysis}
To thoroughly investigate the practicality of ChatGPT 
in the vulnerability detection task (RQ2), we first needed to assess ChatGPT's efficacy in this task. This is because we were interested in exploring how characteristics (e.g., tone) of outputs differ when ChatGPT is accurate versus instances of inaccurate outputs.
To 
this end, we manually analysed the output of all 70 queries 
and provided an assessment as to whether the target SV was detected. This manual assessment was performed independently by the first and third authors of this paper who had a cumulative eight years of both academic and industrial software security experience.

For each SV within the dataset, we used the fixing commit as a baseline for our comparison and employed techniques from prior researchers \cite{croft2023data,10288690} to manually analyse the code changes related to a SV. These techniques are explained step by step as follows. In the first step, we concentrated on the SV summary from the NVD \textit{Description} section. Moreover, we focused on the information available in the source code (i.e., fixing commit). Aside from code changes, the title and message of the fixing commit were also examined, as they usually provide valuable information regarding the functional aspects of the code and the context in which it was altered. Having thoroughly understood the SV, we assessed 
its generated ChatGPT output. 
Accordingly, each output was labelled as either \textit{not found} or \textit{found}. Most often, ChatGPT provided a list of potential vulnerabilities within the input code, with the particular SV we were interested in being mentioned in one of the bullet points. These outputs were classified as \textit{found}. If the output did not contain information about the vulnerability we sought, it would be marked as \textit{not found}. This is even if the other potential vulnerabilities ChatGPT listed appeared genuine. This decision was because we wanted to maximise the reliability of our manual analysis and we did not have the ground truth (i.e., fixing commit) for comparison for the other detected vulnerabilities.

\subsubsection{\textbf{Qualitative Analysis of ChatGPT Outputs}}
\label{sec:output-analysis}
In response to RQ2, we conducted a thematic analysis \cite{cruzes2011recommended} of ChatGPT's outputs regarding the vulnerability detection task. In particular, our primary goal was to demonstrate the breadth and depth of information that a conversational agent such as ChatGPT can offer and examine the nuances of its presentation.
For the thematic analysis of this step, we followed a similar approach as explained in Section \ref{sec:tweet-analysis}. The first and third authors started by collaboratively performing a pilot data labelling of 10 ChatGPT outputs, which constitutes 14\% of the 70 outputs in our dataset. During this process, we focused on the types of information ChatGPT provides and the way in which it is presented. After that, the two authors put more than 100 hours each into this qualitative evaluation and independently analysed the remaining 60 outputs in the dataset. In a supervisory role, the second author reviewed subsets of the analysis results every week to ensure their validity and suggested modifications to improve the assessment accuracy where deemed necessary. Implementing such supervision was necessary to mitigate potential biases and discrepancies in the analysis.
This qualitative analysis was performed with NVivo 12.0 software\cite{nvivo2018} to streamline theme finding. 

\section{Results}
\label{sec:results}

\subsection{\textbf{RQ1: perceptions of using ChatGPT for software security}}
\label{sec:results-RQ1}
We firstly wanted to determine the \textit{occupations} of users of ChatGPT for software security. Such insights provide essential context for the source of discussions and sentiments in our data. Furthermore, we reveal the invested stakeholders of AI-based chatbot solutions for software security. We analysed user profile descriptions of the tweet authors (\textit{n=202}) and identified any occupation details that were listed. We then categorised user occupations using open coding as described in Section \ref{sec:tweet-analysis}. The population of authors was analysed as a set not containing duplicates. This means that a user was only considered once in the population if they authored multiple tweets. We identified five main occupations in our sample data, which we describe below:

\begin{itemize}
    \item \textbf{Security Practitioners} are individuals (\textit{n=78}) that conduct software security or cybersecurity jobs or tasks. This includes occupations such as security analysts, hackers, and vulnerability researchers. 

    \item \textbf{Security Companies} are the Twitter profiles of organisations (\textit{n=27}) that provide software security solutions and services. 

    \item \textbf{Software Practitioners} are individuals (\textit{n=19}) that perform software development or associated tasks for their occupation. 

    \item \textbf{Blogs} are the profiles for online sites or organisations (\textit{n=46}) that provide news outlets or blogging services.

    \item \textbf{Other} defines a collection of occupations which had too few samples to form individual categories. These include academics (\textit{n=6}) and government employees (\textit{n=5}). User profiles that contained insufficient or vague descriptions of occupation (\textit{n=21}) were also placed in the \textbf{Other} category.
\end{itemize}



Table \ref{tab:topics} displays the occupations of Twitter users in our sample. Given that our dataset is comprised of software security discussions, security practitioners are expected to form the largest user demographic (39\%). However, this provides evidence that security practitioners are interested in the potential applications of AI chatbots to their domain, which is supported by the noticeable demographic (13\%) of commercial security companies that are also discussing the viability of such tools. Furthermore, the high percentage of security practitioners in our data indicates that our Twitter sample contains reliable information from users with background domain knowledge. 

We next wanted to examine the \textit{topics} of discussion in our sample Twitter data to understand the areas of software security for which ChatGPT was being used. These insights were expected to help us determine the potential strengths and weaknesses of AI chatbots for different software security tasks. From our qualitative analysis of tweets described in Section \ref{sec:tweet-analysis}, we discovered five main software security categories in which users utilise ChatGPT, which are also displayed in Table \ref{tab:topics} alongside their frequency. We go into detail for each category below. We provide quoted text from the sample tweets that we have analysed. However, many tweets provide associated media to provide further context to the discussion, which we cannot capture within our paper.



\begin{table*}[]
\caption{Software security topics of application for ChatGPT based on tweet authors' occupations.}
\label{tab:topics}
\begin{tabular}{|
>{\columncolor[HTML]{D7CFC6}}c |c|c|c|c|c|
>{\columncolor[HTML]{D7CFC6}}c |}
\hline
\cellcolor[HTML]{ECF4FF}Topic \textbackslash Occupation                                                               & \cellcolor[HTML]{88E1F5}\textbf{Security Practitioner} & \cellcolor[HTML]{88E1F5}\textbf{Security Company} & \cellcolor[HTML]{88E1F5}\textbf{Software Practitioner} & \cellcolor[HTML]{88E1F5}\textbf{Blog} & \cellcolor[HTML]{88E1F5}\textbf{Other} & \textit{\begin{tabular}[c]{@{}c@{}}Topics\\ (Total = 266 Tweets)\end{tabular}} \\ \hline
\textbf{Vulnerability Detection}                                                                                      & 28                                                     & 12                                                & 10                                                     & 19                                    & 9                                      & \textbf{78 (29\%)}                                                             \\
\textbf{Vulnerability Exploits}                                                                                       & 25                                                     & 6                                                 & 2                                                      & 30                                    & 9                                      & \textbf{72 (27\%)}                                                             \\
\textbf{Information Retrieval}                                                                                        & 17                                                     & 6                                                 & 4                                                      & 5                                     & 7                                      & \textbf{39 (15\%)}                                                             \\
\textbf{Code Analysis}                                                                                                & 10                                                     & 6                                                 & 3                                                      & 7                                     & 6                                      & \textbf{32 (12\%)}                                                             \\
\textbf{Other}                                                                                                        & 19                                                     & 6                                                 & 3                                                      & 12                                    & 5                                      & \textbf{45 (17\%)}                                                             \\ \hline
\cellcolor[HTML]{88E1F5}\textit{\begin{tabular}[c]{@{}c@{}}Authors' Occupations\\ (Total = 202 Authors)\end{tabular}} & \cellcolor[HTML]{88E1F5}\textbf{39\%}                  & \cellcolor[HTML]{88E1F5}\textbf{13\%}             & \cellcolor[HTML]{88E1F5}\textbf{9\%}                   & \cellcolor[HTML]{88E1F5}\textbf{23\%} & \cellcolor[HTML]{88E1F5}\textbf{16\%}  & \cellcolor[HTML]{FFFFFF}                                                       \\ \hline
\end{tabular}
\end{table*}

\textit{\textbf{Vulnerability Detection.}} Vulnerability detection is one of the most important components of SV management. Consequently, it is the most frequent discussion category (29\%) within our sample of tweets. Many users were impressed that ChatGPT has even basic capabilities for vulnerability detection due to the difficulty of this task:


\begin{quote}
    \textit{``Experiment time. I fed ChatGPT the Damn Vulnerable Ethereum Smart Contract Migration code in it's entirety. It pointed to four potential vulnerabilities!!!'' (T95)}
\end{quote}

Whilst many users demonstrated that ChatGPT could successfully identify vulnerabilities in simple code snippets, some users were sceptical about the validity of its use in real-world scenarios. 

\begin{quote}
    \textit{``I just went three rounds with a "security researcher" (who was chatGPT in disguise) making reports on our smart contracts. Plausible text and impacts wrapped in magnificent misunderstandings of the basics.'' (T49)}
\end{quote}

ChatGPT produces plausible-sounding answers even for tasks which it cannot necessarily complete successfully. Vulnerability detection requires substantial expertise, so it is hard for many users to verify responses for these tasks. The accuracy of ChatGPT may also be limited as it was not trained to perform these tasks.  

\begin{quote}
    \textit{``Don’t use ChatGPT for security code review. It’s not meant to be used that way, it doesn’t really work (although you might be fooled into thinking it does)'' (T80)}
\end{quote}

However, despite this lack of trust, users still found the vulnerability analysis capabilities helpful in speeding up code review. 

\begin{quote}
    \textit{``I had ChatGPT take code ... and then audit for security. Would I deploy the resulting code? No. But two weeks just became minutes.'' (T150)}
\end{quote}

\textit{\textbf{Vulnerability Exploits.}} Users have also explored leveraging ChatGPT to develop vulnerability exploits for penetration testing or offensive security purposes. These capabilities help security testers verify vulnerabilities, which exceeds the functionality of many traditional static analysis tools.  

\begin{quote}
    \textit{``Write an exploit code that abuse brute-force vulnerability on [One Time Password]. ChatGPT result is mind blowing. Half of the pen tester I know are not even close to that level of well/detailed reporting skills'' (T67)}
\end{quote}

However, a serious concern is that ChatGPT can be similarly used to enable malicious and unethical hacking. 

\begin{quote}
    \textit{``In one instance, a hacker shared an Android malware code written by ChatGPT, which could steal desired files, compress them, and leak them online.'' (T117)}
\end{quote}

\begin{quote}
    \textit{``...ChatGPT can turn anyone into a ransomware and malware threat actor.'' (T163)}
\end{quote}

Such uses would have severe consequences for the state of software security due to the increased ease of cyber attacks and exploits. We observed that discussions about malicious hacking were primarily contained in the \textbf{Speculation} discussion category. Hence, further investigation is required to validate these fears and claims. 

\textit{\textbf{Information Retrieval.}} Users have also used ChatGPT for software security information retrieval and education. 

\begin{quote}
    \textit{``Currently using ChatGPT to help me find and research security solutions for my organization. This has been the biggest game changer for my career since Python.'' (T131)}
\end{quote}

However, similar to vulnerability detection, some security experts were sceptical about the quality of information returned by ChatGPT. A substantial concern is that it is challenging to detect false information provided by ChatGPT.

\begin{quote}
    \textit{``People are excited about using ChatGPT for learning. It's often very good. But the danger is that you can't tell when it's wrong unless you already know the answer. I tried some basic information security questions. In most cases the answers sounded plausible but were in fact BS.'' (T1)}
\end{quote}

\textit{\textbf{Code Analysis.}} ChatGPT can also be used for other secure code analysis tasks, such as understanding security patches, reverse engineering, malware analysis, or decompiling assembly. 

\begin{quote}
    \textit{``...We used ChatGPT to analyse the patch to understand the buffer overflow.'' (T38)}
\end{quote}


\textit{\textbf{Other.}} Finally, users have also theorised the potential utility of ChatGPT for other security tasks, such as threat modelling, vulnerability disclosure, secure configuration, and implementation of security controls. These topics appeared infrequently which implies that did not receive as much attention or interest, but users were still often positive. 

\begin{quote}
    \textit{``... we've not really explored its use-cases for Threat Modeling, which I think is game-changing...'' (T161)}
\end{quote}

\begin{quote}
    \textit{``I had ChatGPT write the breach notification for... Pretty close to the real thing.'' (T94)}
\end{quote}

We finally aimed to investigate the ways in which users were discussing the use of ChatGPT for software security. Our qualitative analysis uncovered five main \textit{types} of discussion. We leveraged these types during our sentiment assessment of the tweets. We assessed user sentiments of ChatGPT for the tweets of each category to better understand the public sentiments about this emerging technology.
\begin{itemize}
    \item \textbf{Article/Tutorial} tweets promote an external article or media source containing information about using ChatGPT for software security. 

    \item \textbf{Speculation} discusses the theoretical use of ChatGPT for software security tasks and topics, with any example usage or evidence of application. 

    \item \textbf{Example} tweets present the authors findings when using example queries as proof of concept application for different software security tasks. Evidence of use is described through experiences or even screenshots of the use cases. 

    \item \textbf{Applied} tweets are similar to \textbf{Example} tweets, except they discuss real-world industrial applications relevant to users' occupations. 
    
    \item \textbf{Tool} tweets promote a tool that has been constructed using ChatGPT technologies. 
\end{itemize}

\begin{figure}[h]
  \centering
  \includegraphics[width=\linewidth]{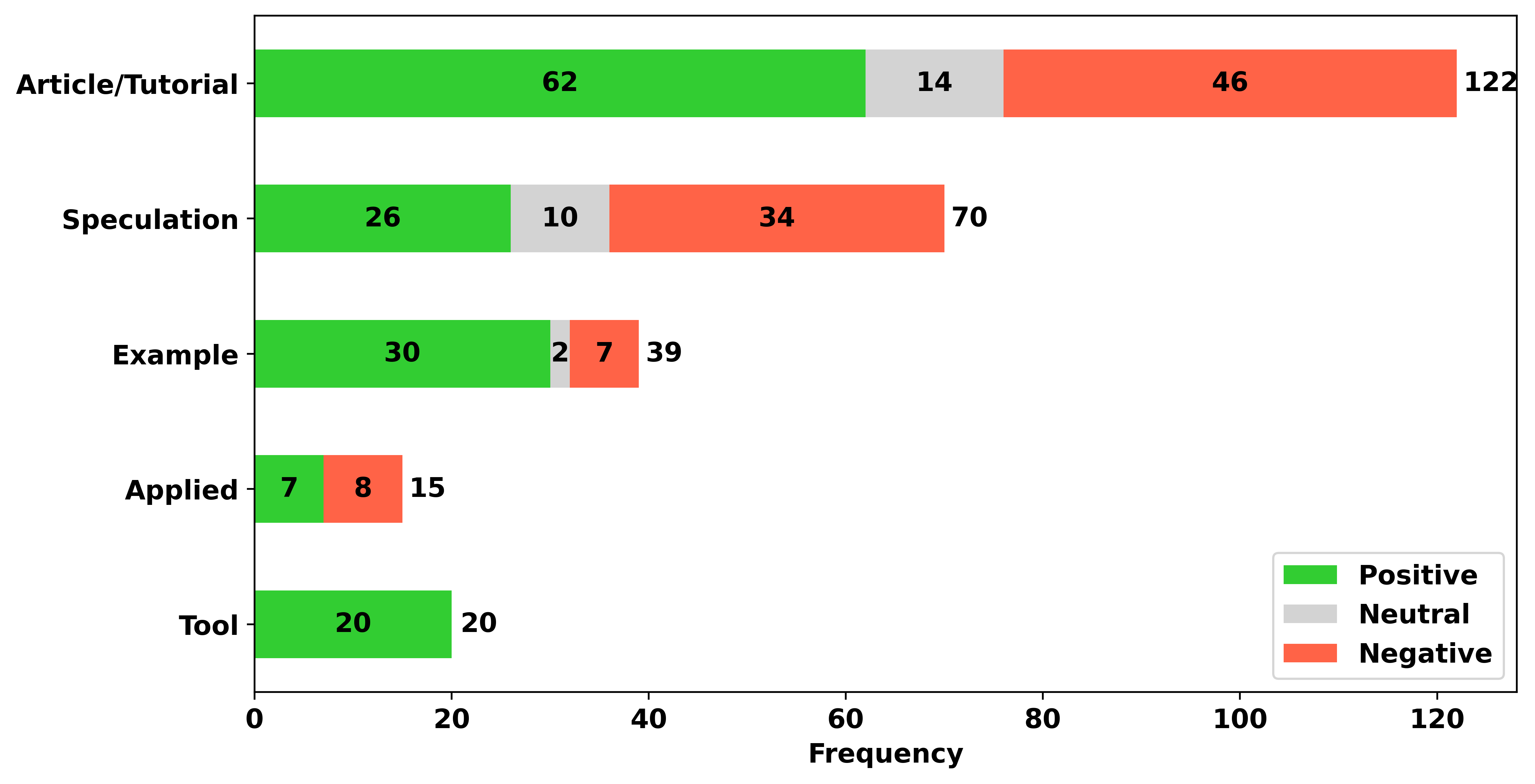}
  \caption{Frequency of discussion types and their sentiments in the Twitter sample data.}
  \label{fig:discussion_sentiment}
\end{figure}

Figure \ref{fig:discussion_sentiment} displays the frequency and user sentiments for each category of discussion.
 Overall sentiments leaned towards positive (54\% positive), which may indicate that most users are optimistic about the capabilities of ChatGPT for software security. 

Articles and tutorials demonstrating how ChatGPT could be used for software security was one of the most prominent discussion types. This demonstrates that the security community is interested in the potential use of AI chatbot technology. Furthermore, some users have begun to develop software security products using ChatGPT or related technologies, as shown in the \textbf{Tool} category. 

Sentiments were also primarily positive for \textbf{Example} tweets (77\%) which indicates that ChatGPT may perform well when used for simple proof-of-concept applications. Similarly, articles that discuss or demonstrate ChatGPT's potential are also slightly positive (51\%). However, applied use has split sentiments, which indicates that ChatGPT may not yet be transferable to real-world software security tasks. 


Furthermore, the sentiment in \textit{speculative} discussions is generally negative (47\%). Hence, many users are sceptical about the actual practical benefits of ChatGPT when used for software security tasks. As the majority of discussion on this topic is either speculation or untested example usage, it is clear that AI chatbots for software security require more investigation and rigorous evaluation. 

\begin{tcolorbox}[right=1pt, left=1pt, top=1pt, bottom=1pt, colback=azure(web)(azuremist)]
\textbf{RQ1 Findings:} 

\begin{itemize}
    
    \item Vulnerability detection was the most discussed software security task among practitioners testing ChatGPT.
    \item The overall sentiment of the discussions among users was positive.
    \item In various security tasks, credibility and practicality of generated information were common concerns.
    \item Sentiments were primarily positive for simple proof-of-concept use cases. 

\end{itemize}
\end{tcolorbox}

\subsection{\textbf{RQ2: practicality of ChatGPT's outputs when utilised for the most desired software security task}}\label{sec:results-rq2}

Using the qualitative analysis described in Section \ref{sec:vuln-analysis}, we first recorded our observations regarding ChatGPT's vulnerability detection accuracy for each SV. 
A total of 43 out of 70 (61.42\%) vulnerabilities were correctly identified by ChatGPT. This efficacy was achieved when ChatGPT was presented with vulnerable code snippets of moderate complexity— specifically, those that are relatively long (average of 32 lines of code per snippet) with focused fixing commits and not overly intricate.

\begin{figure*}[]
  \centering
  \includegraphics[width=0.99\linewidth]{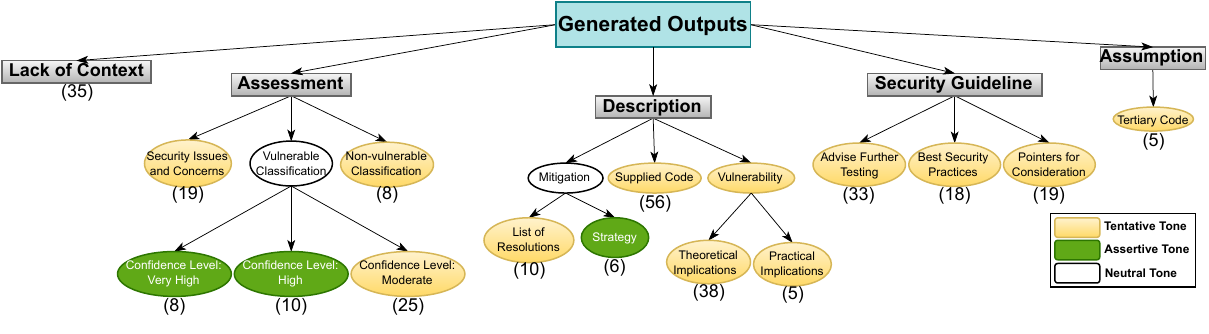}
  \caption{The overall structure of information provided by ChatGPT in vulnerability detection task. Note: the number in parentheses represents the number of instances within each theme}
  \label{fig:output-structure}
\end{figure*}

We answer RQ2 in the following sections by describing the types of information ChatGPT generated and how that information is presented for the vulnerability detection task. This investigation can provide a basis for future enhancements in the design and functionality of tailored conversational agents for software security. We provide quoted text from the analysed ChatGPT outputs. Figure \ref{fig:output-structure} displays the overall structure of the provided information in ChatGPT outputs. 

\subsubsection{\textbf{Assumption}} It has been observed that ChatGPT occasionally makes assumptions about tertiary code when analysing the source code inputted. This kind of information was presented tentatively because they all rest on assumptions.
\begin{quote}
    \textit{``For instance, we're making assumptions about functions like compose\_sadb\_supported() and pfkey\_broadcast() which could have their own set of vulnerabilities.'' (Output \#30)}
\end{quote}


\subsubsection{\textbf{Description}} There was a constant presence of descriptions in the ChatGPT responses. According to our thematic analysis, the descriptions can be divided into three major categories, \textit{Supplied Code}, \textit{Vulnerability}, and \textit{Mitigation}.

\begin{itemize}

    \item \textbf{Supplied Code:} ChatGPT frequently begins its responses with a description of the supplied code. For instance, the functionality of the code or its programming language.
\begin{quote}
    \textit{``This code snippet is written in Java and seems to be responsible for handling requests to trigger a configuration reload through a security token system.'' (Output \#70)}
\end{quote}

    \item \textbf{Vulnerability: }A large portion of each ChatGPT response is devoted to describing identified vulnerabilities. In addition to classifying these descriptions under the \textit{Vulnerability} category, we further divided them into two themes for a more granular analysis: ``Theoretical and Practical Implications". This is because we sought to distinguish between when ChatGPT delves into abstract security implications and when it explains the actionable aspects of potential exploits.

``Theoretical Implication" example:
\begin{quote}
    \textit{``...
4. Exposure of Sensitive Information: The password recovery link may contain sensitive information (like a reset token). The application must make sure to generate a unique and unpredictable token for each password recovery request. ...
'' (Output \#14)}
\end{quote}
``Practical Implication" example:
\begin{quote}
    \textit{``... There are a few places where objects are dereferenced without first ensuring they are not null. This could potentially lead to a segmentation fault and crash the program. ...
'' (Output \#26)}
\end{quote}

Based on our observations, out of 43 cases where ChatGPT identified the specific vulnerability of interest, 38 instances (88\%) were classified under the ``Theoretical Implications'' theme, while only 5 instances  (12\%) were categorised under ``Practical Implications''.

    \item \textbf{Mitigation:} ChatGPT infrequently provided specific information on how to address the particular vulnerability identified. This category contained two themes: ``List of Resolutions" and ``Strategy".
    
    With the ``List of Resolutions" theme, when ChatGPT detects multiple potential vulnerabilities, it offers a numbered list with brief explanations of how to resolve them.
\begin{quote}
    \textit{``To mitigate these vulnerabilities: 
1. Use parameterized queries or prepared statements to avoid SQL injection. 
2. Perform comprehensive input validation and sanitization on all user inputs.
3. Implement proper access control and authorization checks to prevent unauthorized actions.
4. ...
'' (Output \#64)}
\end{quote}
The ``Strategy" theme revealed that ChatGPT provides a 
detailed response 
when encountering a 
severe vulnerability. In addition, we noticed that the responses generated in these instances have an imperative tone. For instance, the phrase \textit{``you should"} is commonly used in this theme.
\begin{quote}
    \textit{``In order to make this script more secure, you should: 
Sanitize and validate all input data to prevent injection attacks. Use a safer file handling technique, which does not allow manipulation of file names. ... '' (Output \#51)}
\end{quote}

\end{itemize}


\subsubsection{\textbf{Security Guideline}} ChatGPT makes a persistent effort to make its responses as comprehensive as possible by including generic security guidelines. This phenomenon was observed in all 70 instances of outputs. We classified these guidelines into three categories: \textit{Advise Further Testing}, \textit{Best Security Practices}, and \textit{Pointers for Consideration}.

\begin{itemize}
    \item \textbf{Advise Further Testing:} The first and most common category (33 instances) is when ChatGPT demands further security audits to be undertaken via experts and automated tools.

\begin{quote}
    \textit{``... It is always recommended to have your
code reviewed by a security expert and use static and dynamic analysis tools to uncover potential security vulnerabilities.'' (Output \#31)}
\end{quote}

    \item \textbf{Best Security Practices:} In the second category, we identified that ChatGPT relied on best security practices (e.g., input validation) applicable to the particular vulnerability to deliver its response.
\begin{quote}
    \textit{``Misconfigurations in PAM can lead to security vulnerabilities, so ensure that you have properly configured PAM in accordance with best security practices. Always adhere to the principle of least privilege. '' (Output \#34)}
\end{quote}

    \item \textbf{Pointers for Consideration:} Lastly, the third category contains very basic security considerations.
\begin{quote}
    \textit{``you might want to consider using a pre-made, widely-reviewed authentication library or service, which will likely have these and other security issues resolved, rather than creating your own.'' (Output \#42)}
\end{quote}
\end{itemize}
Considering that these guidelines were all advice, all three security guideline categories were presented tentatively.

\subsubsection{\textbf{Assessment}} The paramount section of ChatGPT responses pertains to its evaluation of potential vulnerabilities within the submitted source code. Our rigorous analysis identified three assessment categories: \textit{Vulnerable Classification}, \textit{Security Issues and Concerns}, and \textit{Non-vulnerable Classification}.

\begin{itemize}
    \item \textbf{Vulnerable Classification:} This category emerged as the most prominent among the three assessment categories. Considering that all the codes supplied to ChatGPT were already vulnerable, this prevalence was unsurprising. However, a significant variation in the tone of the responses was discovered, which appears to be related directly to the severity of the vulnerability. To further understand, we separated the detected confidence levels into three distinct themes: ``Very High", ``High", and ``Moderate".
    
Within the ``Very High" theme, ChatGPT responses were characterised by an assertive tone. In contrast to the other two themes (High and Moderate), adjectives such as \textit{``significant"} and \textit{``serious"} were frequently used. Furthermore, ChatGPT repeatedly hinted at the presence of multiple vulnerabilities instead of identifying one individually.
\begin{quote}
    \textit{``While this code may seem harmless at first glance, it indeed has several serious security vulnerabilities. ...'' (Output \#9)}
\end{quote}
The tone is less assertive in the ``High" theme compared to the tone in the previous theme. For instance, while ChatGPT continues to identify multiple vulnerabilities, the word \textit{``potential"} is frequently used to indicate a degree of uncertainty.
\begin{quote}
    \textit{``I've found several potential security vulnerabilities in the code. Here's a detailed analysis: ...'' (Output \#53)}
\end{quote}
There is a noticeable drop in certainty at the ``Moderate" theme compared to the ``High" theme. Apart from the frequent use of the term \textit{``potential"} ChatGPT recurrently cites an absence of contextual information as a rationale for its implied uncertainty.
\begin{quote}
    \textit{``Based on the provided code snippet, there are a few potential security vulnerabilities to consider: ... Please remember that a complete and accurate analysis would require more context...'' (Output \#52)}
\end{quote}
Out of all the correctly identified vulnerabilities (43) under investigation, 25 instances (58\%) were classified into this theme. The responses were presented tentatively in this theme compared to the other two themes as ChatGPT was less confident about its outputs.
    \item \textbf{Security Issues and Concerns:} There have been instances where ChatGPT identified potential security weaknesses but refrained from explicitly labeling them as security vulnerabilities. Alternatively, it has used more subtle terms such as \textit{``concerns"} and \textit{``issues"}. We have categorised such instances under the \textit{Security Issues and Concerns} category.
\begin{quote}
    \textit{``However, there are some potential issues to be aware of, more from the perspective of robust programming practices and potential misuse, rather than direct security vulnerabilities: ...'' (Output \#50)}
\end{quote}

    \item \textbf{Non-vulnerable Classification:} ChatGPT made a non-vulnerable assessment in several instances. In these cases, ChatGPT consistently pointed out that the supplied code snippet may not necessarily be secure.
\begin{quote}
    \textit{``At a glance, this code does not seem to contain glaring security vulnerabilities such as buffer overflows, SQL injections, etc., that are commonly associated with insecure code. ...  - an innocent-looking piece of code can become a security hole if used improperly or in an unexpected way.'' (Output \#21)}
\end{quote}
\end{itemize}

\subsubsection{\textbf{Lack of Context}} 
We observed that in exactly \%50 of the outputs (35 out of 70), ChatGPT mentioned that more context would be required for a concrete security analysis.
\begin{quote}
    \textit{``While the code snippet does not provide all of the details, we can still make several general observations based on the provided information.'' (Output \#13)}
\end{quote}

\begin{tcolorbox}[right=1pt, left=1pt, top=1pt, bottom=1pt, colback=azure(web)(azuremist)]
\textbf{RQ2 Findings:} 

\begin{itemize}
    \item Even when ChatGPT correctly pinpoints the vulnerability of interest, it predominantly focuses on abstract rather than practical security implications.
    \item Although ChatGPT responses appear comprehensive, they are heavily populated with generic security information (e.g., Description, Guideline).
    \item The severity of the vulnerability appears to be directly related to the tone of the responses.
    \item ChatGPT generally adopted a tentative tone to present the information.
    
\end{itemize}
\end{tcolorbox}

\section{Discussion}
\label{sec:discussion}
In RQ1, our investigation has shown that software and security practitioners enthusiastically experiment with and share the pros/cons of using ChatGPT for assistance in various software security tasks, such as vulnerability detection and security information retrieval. However, the state of discussions mostly remains in example proof of concept use, whereas only a few cases of applied use have been discussed. Current security practitioners appear to think that ChatGPT is inaccurate as a standalone solution. Hence, ChatGPT may be better suited to secondary tasks such as vulnerability report writing. For instance, users can draft comprehensive and detailed vulnerability reports by providing context-specific prompts and formatting instructions. Nevertheless, security professionals should review and validate the contents to ensure accuracy and completeness.

In the second phase of our study (RQ2), we treated ChatGPT as an oracle for the vulnerability detection task. In our data sample, it achieved a detection accuracy of 61\%, which is far from ideal. Furthermore, after qualitatively examining the generated responses, we determined the characteristics (i.e., type and presentation style of information) of this chatbot and found that it is unable to provide reliable information for this task. Our analysis revealed that ChatGPT responses suffer from two specific quality issues. The most prevalent issue we observed was that all responses (70 instances) contained generic security information, as indicated in the ``\textit{Theoretical Implication}'' theme and the three themes under the ``\textit{Security Guideline}'' information type. Throughout these themes, we found an abundance of unneeded information, resulting in a lack of clarity in the responses. This issue persisted even in cases where ChatGPT correctly identified the vulnerability. Another noteworthy issue we found was that ChatGPT regularly lacks assertiveness when delivering responses. Based on our analysis, the majority of responses were given with a tentative tone, as shown in Figure \ref{fig:output-structure}, and a consistent request for additional contextual information. 
We believe both of these quality issues may undermine the practicality of ChatGPT for vulnerability detection within industry settings. This is because when a fault localisation tool is not transparent about its limitations, developers may inadvertently spend valuable time and effort verifying potentially inaccurate outputs \cite{le2015should}.

However, in contrast to existing weakness detection tools (e.g., static code analyser), conversational agents such as ChatGPT allow users to interface with a tool with profound natural language comprehension abilities. This potentially can have positive effects on novice learners who want to start learning secure coding principles. 

Overall, we advocate the need for further development of domain-relevant LLMs trained using software security data. Although ChatGPT is trained for general use and not intended for software security, our qualitative study shows it can assist and augment processes to a certain degree. Hence, we expect domain-specific LLMs to have potent capabilities for software security in the future. These technologies ultimately need to undergo rigorous industrial evaluation for which the empirical software engineering community can provide leadership in collaboration with AI researchers and practitioners. 

\subsection{\textbf{Implications of Our Study}}
\textbf{Programmers.} Understanding the nuances of responses (RQ2) generated by ChatGPT can offer valuable benefits to programmers who leverage this technology for vulnerability detection. Notably, they may better assess the legitimacy of the generated information and make informed decisions about whether to trust and adopt the security suggestions. Furthermore, by understanding ChatGPT's strengths and limitations, programmers can better integrate this technology into their development workflows, using it as an adjunct to other weakness detection tools. 

\textbf{AI practitioners.} The investigation into ChatGPT's communication of software security information (RQ2) has explicit implications for AI researchers and developers. Insights of this nature may assist them in fine-tuning LLM-based chatbots for specific software security tasks. Specifically, they must strive to create security-tailored LLMs capable of conveying security information to users in a manner that is more straightforward, actionable, and less susceptible to misinterpretation. Such models are likely to perform better and be more practical when applied in real-world security settings.

\textbf{Security researchers.} We have identified some promising research directions for researchers. Our analysis in RQ1 unveiled that there is a lot of speculation (discussions in the Vulnerability Exploits category) regarding the potential misuse of ChatGPT for malicious purposes. Researchers need to identify whether technologies like ChatGPT can enable malicious hacking so that countermeasures can be applied to prevent using these technologies for such purposes. 
Additionally, our analysis in RQ2  demonstrated that ChatGPT alters its language tone when confronted with vulnerabilities of varying severity scores. Further research is required to thoroughly understand the capability of LLMs in estimating vulnerability severity. This capability is imperative in vulnerability prioritisation, where precise severity assessment is essential for effective mitigation strategies.



\section{Threats to Validity}
\label{sec:threats}

\textit{Construct validity:} Our manual analysis may be subject to bias or inaccuracies, as is the case for most qualitative research. We have attempted to minimise such threats to validity by ensuring that two annotators independently performed the required manual analysis.

\textit{Internal Validity:} We acknowledge that the criteria we used for selecting vulnerability data could have influenced the outcome of our study. As explained in Section \ref{sec:dataset-vulnerability}, our intention to manually trace code changes meant we could not trace tangled fixing commits. As a result, we decided to limit the number of line changes within fixing commits to 10 lines. This approach ensured we gathered code snippets that were neither simplistic nor too complex.

\textit{External Validity:} Qualitative analysis enables us to examine our data in depth, but consequently, we can only engage with a small sample of the total dataset. We cannot generalise beyond the sampled tweets and vulnerabilities due to the effort required to examine the data manually. We obtained our tweet sample based on the most liked and influential tweets, so we expect our sample represents a larger audience. 

Furthermore, public opinions may be inaccurate or biased for emerging technologies, as users may be overly excited or not have had enough time to evaluate their use properly. Despite our desire to extend the three-month period for collecting tweets, we were unable to do so since Twitter's academia API has been deprecated following recent changes at the company.

It is also possible that our findings might not generalise to future ChatGPT versions enriched with updated data. However, we conducted this study using a single ChatGPT version with the same knowledge cut-off date (Sep 2021) for consistency.

\section{Conclusion}
\label{sec:conclusion}


We have conducted an empirical investigation to understand the potential use of ChatGPT for software security. Based on examining public discussions regarding this LLM, we found that practitioners were often optimistic about its utility for various security tasks. 
Our systematic evaluation further demonstrated that ChatGPT can identify vulnerabilities 
with roughly 61\% accuracy. Furthermore, we qualitatively demonstrated the versatility and granularity of information that ChatGPT can offer in the vulnerability detection task. The results revealed that ChatGPT outputs in this task are often too generic and ambiguous to be of practical use at the industry level. A tailored security LLM could enhance this capability, enabling practitioners to leverage the abundance of knowledge embedded in LLMs.

\section{Data Availability}
We have made all our artefacts of this study available via a reproduction package\cite{reproduction_package}.

\section*{Acknowledgment}
This work has been supported by the Cyber Security Cooperative Research Centre Limited whose activities are partially funded by the Australian Government’s Cooperative Research Centre Programme.

\bibliographystyle{IEEEtran}
\bibliography{bibfile}

\begin{thebibliography}{10}
\providecommand{\url}[1]{#1}
\csname url@samestyle\endcsname
\providecommand{\newblock}{\relax}
\providecommand{\bibinfo}[2]{#2}
\providecommand{\BIBentrySTDinterwordspacing}{\spaceskip=0pt\relax}
\providecommand{\BIBentryALTinterwordstretchfactor}{4}
\providecommand{\BIBentryALTinterwordspacing}{\spaceskip=\fontdimen2\font plus
\BIBentryALTinterwordstretchfactor\fontdimen3\font minus \fontdimen4\font\relax}
\providecommand{\BIBforeignlanguage}[2]{{%
\expandafter\ifx\csname l@#1\endcsname\relax
\typeout{** WARNING: IEEEtran.bst: No hyphenation pattern has been}%
\typeout{** loaded for the language `#1'. Using the pattern for}%
\typeout{** the default language instead.}%
\else
\language=\csname l@#1\endcsname
\fi
#2}}
\providecommand{\BIBdecl}{\relax}
\BIBdecl

\bibitem{rajapakse2022challenges}
R.~N. Rajapakse, M.~Zahedi, M.~A. Babar, and H.~Shen, ``Challenges and solutions when adopting devsecops: A systematic review,'' \emph{Information and Software Technology}, vol. 141, p. 106700, 2022.

\bibitem{van2020schrodinger}
D.~Van Der~Linden, P.~Anthonysamy, B.~Nuseibeh, T.~T. Tun, M.~Petre, M.~Levine, J.~Towse, and A.~Rashid, ``Schr{\"o}dinger's security: Opening the box on app developers' security rationale,'' in \emph{2020 IEEE/ACM 42nd International Conference on Software Engineering (ICSE)}.\hskip 1em plus 0.5em minus 0.4em\relax IEEE, 2020, pp. 149--160.

\bibitem{mousavi2023detecting}
Z.~Mousavi, C.~Islam, M.~A. Babar, A.~Abuadbba, and K.~Moore, ``Detecting misuses of security apis: A systematic review,'' \emph{arXiv preprint arXiv:2306.08869}, 2023.

\bibitem{mcgraw2008automated}
G.~McGraw, ``Automated code review tools for security,'' \emph{Computer}, vol.~41, no.~12, pp. 108--111, 2008.

\bibitem{frazier2022investigating}
M.~Frazier, S.~Kumar, K.~Damevski, and L.~Pollock, ``Investigating user perceptions of conversational agents for software-related exploratory web search,'' in \emph{Proceedings of the ACM/IEEE 44th International Conference on Software Engineering: New Ideas and Emerging Results}, 2022, pp. 51--55.

\bibitem{adamopoulou2020overview}
E.~Adamopoulou and L.~Moussiades, ``An overview of chatbot technology,'' in \emph{IFIP International Conference on Artificial Intelligence Applications and Innovations}.\hskip 1em plus 0.5em minus 0.4em\relax Springer, 2020, pp. 373--383.

\bibitem{erlenhov2020empirical}
L.~Erlenhov, F.~G. D.~O. Neto, and P.~Leitner, ``An empirical study of bots in software development: Characteristics and challenges from a practitioner’s perspective,'' in \emph{Proceedings of the 28th ACM joint meeting on european software engineering conference and symposium on the foundations of software engineering}, 2020, pp. 445--455.

\bibitem{tony2022conversational}
C.~Tony, M.~Balasubramanian, N.~E. D{\'\i}az~Ferreyra, and R.~Scandariato, ``Conversational devbots for secure programming: An empirical study on skf chatbot,'' in \emph{Proceedings of the International Conference on Evaluation and Assessment in Software Engineering 2022}, 2022, pp. 276--281.

\bibitem{brown2020language}
T.~Brown, B.~Mann, N.~Ryder, M.~Subbiah, J.~D. Kaplan, P.~Dhariwal, A.~Neelakantan, P.~Shyam, G.~Sastry, A.~Askell \emph{et~al.}, ``Language models are few-shot learners,'' \emph{Advances in neural information processing systems}, vol.~33, pp. 1877--1901, 2020.

\bibitem{haque2022think}
M.~U. Haque, I.~Dharmadasa, Z.~T. Sworna, R.~N. Rajapakse, and H.~Ahmad, ``" i think this is the most disruptive technology": Exploring sentiments of chatgpt early adopters using twitter data,'' \emph{arXiv preprint arXiv:2212.05856}, 2022.

\bibitem{openai2023gpt4}
OpenAI, ``Gpt-4 technical report,'' 2023.

\bibitem{thapa2022transformer}
C.~Thapa, S.~I. Jang, M.~E. Ahmed, S.~Camtepe, J.~Pieprzyk, and S.~Nepal, ``Transformer-based language models for software vulnerability detection,'' in \emph{Proceedings of the 38th Annual Computer Security Applications Conference}, 2022, pp. 481--496.

\bibitem{liu2023no}
Z.~Liu, Y.~Tang, X.~Luo, Y.~Zhou, and L.~F. Zhang, ``No need to lift a finger anymore? assessing the quality of code generation by chatgpt,'' \emph{arXiv preprint arXiv:2308.04838}, 2023.

\bibitem{sobania2023analysis}
D.~Sobania, M.~Briesch, C.~Hanna, and J.~Petke, ``An analysis of the automatic bug fixing performance of chatgpt,'' \emph{arXiv preprint arXiv:2301.08653}, 2023.

\bibitem{fu2023chatgpt}
M.~Fu, C.~Tantithamthavorn, V.~Nguyen, and T.~Le, ``Chatgpt for vulnerability detection, classification, and repair: How far are we?'' \emph{arXiv preprint arXiv:2310.09810}, 2023.

\bibitem{tian2017apibot}
Y.~Tian, F.~Thung, A.~Sharma, and D.~Lo, ``Apibot: question answering bot for api documentation,'' in \emph{2017 32nd IEEE/ACM international conference on automated software engineering (ASE)}.\hskip 1em plus 0.5em minus 0.4em\relax IEEE, 2017, pp. 153--158.

\bibitem{kehagiasempirical}
I.~Kalouptsoglou, M.~Siavvas, A.~Ampatzoglou, D.~Kehagias, and A.~Chatzigeorgiou, ``An empirical comparison of transformer-based models in vulnerability prediction,'' 05 2023.

\bibitem{chan2023transformer}
A.~Chan, A.~Kharkar, R.~Z. Moghaddam, Y.~Mohylevskyy, A.~Helyar, E.~Kamal, M.~Elkamhawy, and N.~Sundaresan, ``Transformer-based vulnerability detection in code at edittime: Zero-shot, few-shot, or fine-tuning?'' \emph{arXiv preprint arXiv:2306.01754}, 2023.

\bibitem{chen2023chatgpt}
C.~Chen, J.~Su, J.~Chen, Y.~Wang, T.~Bi, Y.~Wang, X.~Lin, T.~Chen, and Z.~Zheng, ``When chatgpt meets smart contract vulnerability detection: How far are we?'' \emph{arXiv preprint arXiv:2309.05520}, 2023.

\bibitem{szabo2023new}
Z.~Szab{\'o} and V.~Bilicki, ``A new approach to web application security: Utilizing gpt language models for source code inspection,'' \emph{Future Internet}, vol.~15, no.~10, p. 326, 2023.

\bibitem{ozturk2023new}
O.~S. Ozturk, E.~Ekmekcioglu, O.~Cetin, B.~Arief, and J.~Hernandez-Castro, ``New tricks to old codes: can ai chatbots replace static code analysis tools?'' in \emph{Proceedings of the 2023 European Interdisciplinary Cybersecurity Conference}, 2023, pp. 13--18.

\bibitem{zhang2023prompt}
C.~Zhang, H.~Liu, J.~Zeng, K.~Yang, Y.~Li, and H.~Li, ``Prompt-enhanced software vulnerability detection using chatgpt,'' \emph{arXiv preprint arXiv:2308.12697}, 2023.

\bibitem{feng2020codebert}
Z.~Feng, D.~Guo, D.~Tang, N.~Duan, X.~Feng, M.~Gong, L.~Shou, B.~Qin, T.~Liu, D.~Jiang \emph{et~al.}, ``Codebert: A pre-trained model for programming and natural languages,'' \emph{arXiv preprint arXiv:2002.08155}, 2020.

\bibitem{ullah2023can}
S.~Ullah, M.~Han, S.~Pujar, H.~Pearce, A.~Coskun, and G.~Stringhini, ``Can large language models identify and reason about security vulnerabilities? not yet,'' \emph{arXiv preprint arXiv:2312.12575}, 2023.

\bibitem{nong2024chain}
Y.~Nong, M.~Aldeen, L.~Cheng, H.~Hu, F.~Chen, and H.~Cai, ``Chain-of-thought prompting of large language models for discovering and fixing software vulnerabilities,'' \emph{arXiv preprint arXiv:2402.17230}, 2024.

\bibitem{mathews2024llbezpeky}
N.~S. Mathews, Y.~Brus, Y.~Aafer, M.~Nagappan, and S.~McIntosh, ``Llbezpeky: Leveraging large language models for vulnerability detection,'' \emph{arXiv preprint arXiv:2401.01269}, 2024.

\bibitem{croft2023data}
R.~Croft, M.~A. Babar, and M.~M. Kholoosi, ``Data quality for software vulnerability datasets,'' in \emph{2023 IEEE/ACM 45th International Conference on Software Engineering (ICSE)}.\hskip 1em plus 0.5em minus 0.4em\relax IEEE, 2023, pp. 121--133.

\bibitem{gennari2024considerations}
J.~Gennari, S.-h. Lau, S.~Perl, J.~Parish, and G.~Sastry, ``Considerations for evaluating large language models for cybersecurity tasks,'' 2024.

\bibitem{sallou2023breaking}
J.~Sallou, T.~Durieux, and A.~Panichella, ``Breaking the silence: the threats of using llms in software engineering,'' \emph{arXiv preprint arXiv:2312.08055}, 2023.

\bibitem{singer2014software}
L.~Singer, F.~Figueira~Filho, and M.-A. Storey, ``Software engineering at the speed of light: how developers stay current using twitter,'' in \emph{Proceedings of the 36th International Conference on Software Engineering}, 2014, pp. 211--221.

\bibitem{dong2021review}
X.~Dong and Y.~Lian, ``A review of social media-based public opinion analyses: Challenges and recommendations,'' \emph{Technology in Society}, vol.~67, p. 101724, 2021.

\bibitem{bian2016mining}
J.~Bian, K.~Yoshigoe, A.~Hicks, J.~Yuan, Z.~He, M.~Xie, Y.~Guo, M.~Prosperi, R.~Salloum, and F.~Modave, ``Mining twitter to assess the public perception of the “internet of things”,'' \emph{PloS one}, vol.~11, no.~7, p. e0158450, 2016.

\bibitem{NVD}
\BIBentryALTinterwordspacing
NIST, ``\BIBforeignlanguage{en}{National vulnerability database}.'' [Online]. Available: \url{https://nvd.nist.gov/}
\BIBentrySTDinterwordspacing

\bibitem{cochran2007sampling}
W.~G. Cochran, \emph{Sampling techniques}.\hskip 1em plus 0.5em minus 0.4em\relax John Wiley \& Sons, 2007.

\bibitem{le2024latent}
T.~H. Le, X.~Du, and M.~A. Babar, ``Are latent vulnerabilities hidden gems for software vulnerability prediction? an empirical study,'' \emph{arXiv preprint arXiv:2401.11105}, 2024.

\bibitem{CWE}
\BIBentryALTinterwordspacing
Mitre, ``\BIBforeignlanguage{en}{Common weakness enumeration}.'' [Online]. Available: \url{https://cwe.mitre.org/}
\BIBentrySTDinterwordspacing

\bibitem{white2023chatgpt}
J.~White, S.~Hays, Q.~Fu, J.~Spencer-Smith, and D.~C. Schmidt, ``Chatgpt prompt patterns for improving code quality, refactoring, requirements elicitation, and software design,'' \emph{arXiv preprint arXiv:2303.07839}, 2023.

\bibitem{feng2023prompting}
S.~Feng and C.~Chen, ``Prompting is all your need: Automated android bug replay with large language models,'' \emph{arXiv preprint arXiv:2306.01987}, 2023.

\bibitem{cruzes2011recommended}
D.~S. Cruzes and T.~Dyba, ``Recommended steps for thematic synthesis in software engineering,'' in \emph{2011 international symposium on empirical software engineering and measurement}.\hskip 1em plus 0.5em minus 0.4em\relax IEEE, 2011, pp. 275--284.

\bibitem{cohen1960}
J.~Cohen, ``A coefficient of agreement for nominal scales,'' \emph{Educational and psychological measurement}, vol.~20, no.~1, pp. 37--46, 1960.

\bibitem{bird2009natural}
S.~Bird, E.~Klein, and E.~Loper, \emph{Natural language processing with Python: analyzing text with the natural language toolkit}.\hskip 1em plus 0.5em minus 0.4em\relax " O'Reilly Media, Inc.", 2009.

\bibitem{10288690}
M.~M. Kholoosi, M.~A. Babar, and C.~Yilmaz, ``Empirical analysis of software vulnerabilities causing timing side channels,'' in \emph{2023 IEEE Conference on Communications and Network Security (CNS)}, 2023, pp. 1--9.

\bibitem{nvivo2018}
\BIBentryALTinterwordspacing
Q.~I.~P. Ltd., ``Nvivo,'' 2018. [Online]. Available: \url{https://www.qsrinternational.com/nvivo-qualitative-data-analysis-software/home}
\BIBentrySTDinterwordspacing

\bibitem{le2015should}
T.-D.~B. Le, D.~Lo, and F.~Thung, ``Should i follow this fault localization tool’s output? automated prediction of fault localization effectiveness,'' \emph{Empirical Software Engineering}, vol.~20, pp. 1237--1274, 2015.

\bibitem{reproduction_package}
\BIBentryALTinterwordspacing
M.~Kholoosi, A.~Babar, and R.~Croft, ``Reproduction package,'' 8 2024. [Online]. Available: \url{https://figshare.com/articles/dataset/Reproduction_package_for_paper_A_Qualitative_Study_on_Using_ChatGPT_for_Software_Security_Perception_vs_Practicality_/24452365}
\BIBentrySTDinterwordspacing

\end{thebibliography}

\end{document}